\documentstyle[12pt,fleqn]{article}
\newcommand{\eeq}{\end{eqnarray}}
\newcommand{\beq}{\begin{eqnarray}}
\newcommand{\bD}{{\bf D}}
\newcommand{\bE}{{\bf E}}
\newcommand{\tD}{\widetilde{\bf D}}
\newcommand{\tE}{\widetilde{\bf E}}
\newcommand{\tph}{\widetilde{\phi}}

\date{}
\title{\bf Generation of  multipole moments by external field
in Born-Infeld non-linear electrodynamics}
\author{\\ \\ Dariusz Chru\'sci\'nski\footnotemark \\
       Fakult\"at f\"ur Physik, Universit\"at Freiburg\\
       Hermann-Herder-Str. 3, D-79104 Freiburg, Germany\\ \\ \\
Jerzy Kijowski\\ Centrum Fizyki Teoretycznej PAN\\
        Aleja Lotnik\'ow 32/46, 02-668 Warsaw, Poland}
\begin{document}

\def\thefootnote{\relax}\footnotetext{$^*$On leave from
 Institute of Physics, Nicholas Copernicus
University, ul. Grudzi\c{a}dzka 5/7, 87-100 Toru\'n, Poland.}

\maketitle
\begin{abstract}

The mechanism for generation of  multipole moments due to an external field
is presented for the Born-Infeld charged particle. The ``polarizability
coefficient'' $\kappa_l$ for arbitrary $l$-pole moment is calculated. It
turns out, that $\kappa_l
\sim r_0^{2l+1}$, where
$r_0 := \sqrt{|e| / 4\pi b}\ $ and $b$ is the Born-Infeld non-linearity
constant. Physical implications are considered.

\end{abstract}

\vspace{4cm} \noindent Freiburg THEP-97/32

\newpage

\section{Introduction}

Recently, one of us has proposed a consistent, relativistic theory of
the classical Maxwell field interacting with classical, charged,
point-like particles (cf. \cite{EMP}). For this purpose an ``already
renormalized'' formula for the total four-momentum of a system composed
of both the moving particles and the surrounding electromagnetic field
was used.  It was proved, that the conservation of the total
four-momentum defined by this formula is equivalent to a certain
boundary condition for the behaviour of the Maxwell field in the
vicinity of the particle trajectories. Without this condition, Maxwell
theory with point-like sources is not dynamically closed: initial
conditions for particles and fields do not imply uniquely   the future
and the past of the system. Indeed, the particles
 trajectories fulfilling initial conditions can be {\em arbitrarily}
chosen and then the initial-value-problem for the field alone can be 
solved uniquely.
The boundary condition derived this way was called 
{\it fundamental equation}. When added to Maxwell
equations, it provides the missing dynamical equation: now, particles
trajectories cannot be chosen arbitrarily and initial data uniquely
imply the future and the past evolution of the composed ``particles +
fields'' system.

Physically, the ``already renormalized'' formula for the total
four-mo\-men\-tum was suggested by a suitable approximation procedure
applied to an extended-particle model. In such a model we suppose that
the particle is a stable, soliton-like solution of a hypothetical
fundamental theory of interacting electromagnetic and matter fields.
We assume that this hypothetical theory tends asymptotically to the
linear Maxwell electrodynamics, in the weak field regime (i.~e.~for
weak electromagnetic fields and ``almost vanishing'' matter fields).
This means, that ``outside of the particles'' the entire theory reduces
to Maxwell electrodynamics. Starting from this model, a formula was
found, which gives in a good approximation the total four-momentum of
the system composed of both the moving (extended) particles and the
surrounding electromagnetic field.  This formula uses only the
``mechanical'' information about the particle (position, velocity, mass
$m$ and the electric charge $e$) and the free electromagnetic field
outside of the particle. It turns out, that this formula does not
produce any infinities when applied to the case of point particles,
i.~e.~it is ``already renormalized''.  Using this philosophy, this
formula was taken as a starting
point for a mathematically self-consistent theory of {\it point-like
particles} interacting with the linear Maxwell field. The ``fundamental
equation'' of the theory is precisely the conservation of the total
four momentum of the system ``particles + fields'', defined by this
formula.

At this point a natural idea arises, to construct a
``second-generation'' theory, which approximates better the real
properties of an extended particle, and takes into account also
possible deformations of its interior, due to strong external field. In
\cite{dipole}  a simple mechanism of generation of a
particle's electric dipole moment  was proposed. As a specific
particle's model 
 we have used the Born-Infeld particle described by 
 the Born-Infeld non-linear electrodynamics.

In the present paper we prove that a similar mechanism is responsible 
for the generation of higher multipole moments.

Mathematically,   such a polarizability is due to the elliptic properties
of the field equations describing the statics of the physical system 
under consideration.  Given a particular model of the matter fields
interacting with electromagnetism, the ``particle at rest''-solution
corresponds to a minimum of the total field energy. It is, therefore,
described by a solution of a system of elliptic equations
(Euler-Lagrange equations derived from the total Hamiltonian of the
hypothetical fundamental theory of interacting matter fields and
electromagnetic field). Far away from the particle, these equations
reduce to the free Maxwell equations.

This solution corresponds to the vanishing
boundary conditions at infinity.  Physical situation ``particle in a
non-vanishing external field'' corresponds to the solution of 
{\it the same} elliptic equations but with  {\em non-vanishing} boundary 
conditions ${\bf E}_{\infty}$ at infinity. More precisely, we assume 
that the electric field at infinity behaves like
\beq  
E^k({\bf x}) = E^k_{reg}({\bf x}) + E^k_{\infty}({\bf x})\ ,
\eeq
where
\beq   \label{E-infty}
E^k_{\infty}({\bf x}) := {\cal Q}^k_{\ i_2...i_l}x^{i_2}...x^{i_l}\ ,
\eeq
(the $l$-pole tensor ${\cal Q}_{i_1i_2...i_l}$ is completely
 symmetric and traceless) and the regular part $E^k_{reg}({\bf x})$ 
vanishes at infinity.

Suppose that the free particle (the unperturbed solution) displays no
internal structure. This means that for vanishing external field
 ${\cal Q}=0$ the regular part $E^k_{reg}$ reduces to the simplest 
Coulomb field describing the monopole with a given electric charge $e$.
But for nontrivial perturbation $\cal Q$ the regular field may contain
an extra multipole term at infinity, of the type:
\beq    \label{E-M}
E^k_{{\cal M}} := \frac{ r^2 {\cal M}^k_{\ i_2...i_l}x^{i_2}...x^{i_l}
- \frac{2l+1}{l}x^k {\cal M}_{i_1i_2...i_l}x^{i_1}x^{i_2}...x^{i_l} }
{ r^{3+2l} } \ .
\eeq
The reason for  creation of this extra multipole moments is the field 
nonlinearity in the strong field region.
 For weak perturbations, the
relation between the particle's $l$-pole moment ${\cal M}_{i_1i_2...i_l}$
created this way and the $l$-pole moment ${\cal Q}_{i_1i_2...i_l}$ of the
external  field  is expected to be linear in the first
approximation:
\begin{eqnarray}   \label{M-Q}
{\cal M}_{i_1i_2...i_l} = \kappa_l \, {\cal Q}_{i_1i_2...i_l} \
\end{eqnarray}
 and the coefficient $\kappa_l$ describes the ``deformability''
of the particle, due to non-linear character of the interaction of the matter
fields (constituents of the particle) with the electromagnetic field.

The coefficients $\kappa_l$ arise, therefore, similarly as ``reflection''
or ``transmission'' coefficients in the scattering theory. Formulae 
(\ref{E-infty})  and (\ref{E-M}) describe two independent solutions of the 
second order, linear, elliptic equation describing the free, statical 
Maxwell field surrounding the particle.
 Outside of the particle they may be mixed in
an arbitrary proportion. Such an arbitrary mixture is no longer
possible if it has to match an exact solution of non-linear equations
describing the interior of the particle. The relation (\ref{M-Q})
arises, therefore, as the maching condition between these two solutions.

In the present paper we assume that  the unperturbed
particle is described by the spherically symmetric, static solution of
non-linear Born-Infeld electrodynamics with a $\delta$-like source.
 We find explicitly the
two-dimensional family of all the $l$-pole perturbations of the
above solution.  They all behave correctly at $r \rightarrow \infty$.
For $r \rightarrow 0$, however, there is one perturbation which remains
regular, and another one which increases faster then the unperturbed
solution. The variation of the total field energy due to the latter
perturbation is divergent in the vicinity of the particle, 
which we consider to be an unphysical feature.
We conclude that all the physically admissible perturbations must be 
proportional to the one which is regular at $0$. At $r \rightarrow
\infty$ this solution behaves as a mixture of solutions (\ref{E-infty})
and (\ref{E-M}). We calculate the ratio
between these two ingredients and we interpret it as the
$l$-th polarizability coefficient of the Born-Infeld particle.

\section{Perturbations of the Born-Infeld particle}

The Born-Infeld electrodynamics \cite{BI} (see also \cite{IBB}) is defined
by the following lagrangian
\begin{eqnarray}    \label{Lag-BI}
{\cal L}_{BI} := b^2\left[1- \sqrt{1-2b^{-2}S - b^{-4}P^2}\right]\ ,
\end{eqnarray}
where $S$ and $P$ are the following Lorentz invariants:
\begin{eqnarray}
S &:=& -\frac 14 f_{\mu\nu}f^{\mu\nu}\ = \frac 12 ({\bf E}^2-{\bf B}^2)\
,\\ P &:=& -\frac 14
\epsilon_{\mu\nu\lambda\kappa}f^{\mu\nu}f^{\lambda\kappa}\ = {\bf
E}{\bf B}\ ,
\end{eqnarray}
and $f_{\mu\nu}$ is a tensor of the electromagnetic field defined in a
standard way {\it via} a four-potential vector. The parameter ``$b$''  has
a dimension of a field strength (Born and Infeld called it the {\it
absolute field}, cf. \cite{BI}) and it measures the nonlinearity of the
theory. In the limit $b \rightarrow \infty$ the lagrangian ${\cal L}_{BI}$
tends to the standard Maxwell lagrangian
\begin{eqnarray}     \label{Lag-Max}
{\cal L}_{Maxwell} = S\ .
\end{eqnarray}
Note, that  field equations derived from (\ref{Lag-BI}) have the same form
as Maxwell equations derived from (\ref{Lag-Max}), i.~e.
\begin{eqnarray}       \label{Max1}
\nabla \times {\bf E} + \dot{{\bf B}} &=& 0\ ,\ \ \ \ \nabla \cdot{\bf B}=0\
,\\
\nabla \times {\bf H} - \dot{{\bf D}} &=& {\bf j}\ ,\ \ \ \ \nabla\cdot{\bf D}
 =\rho\ ,         \label{Max2}
\end{eqnarray}
(to obtain the Born-Infeld equations with sources $\rho$ and {\bf j} one
has to add to ${\cal L}_{BI}$ the standard interaction lagrangian ``$j^\mu
A_\mu$'').  However, the relation between fields ({\bf E},{\bf B}) and
({\bf D},{\bf H}) is now highly nonlinear:
\begin{eqnarray}
{\bf D} &:=& \frac{\partial {\cal L}_{BI}}{\partial {\bf E}} =
\frac{ {\bf E} + b^{-2}({\bf E}{\bf B}){\bf B} }
{\sqrt{ 1 - b^{-2}({\bf E}^2 - {\bf B}^2) - b^{-4}({\bf E}{\bf B})^2 }}\ ,
\label{D} \\ {\bf H} &:=& -\frac{\partial {\cal L}_{BI}}{\partial {\bf B}}
=
\frac{ {\bf B} - b^{-2}({\bf E}{\bf B}){\bf E} }
{\sqrt{ 1 - b^{-2}({\bf E}^2 - {\bf B}^2) - b^{-4}({\bf E}{\bf B})^2 }}\ .
\label{H}
\end{eqnarray}
The above formulae are responsible for the nonlinear character of the
Born-Infeld theory. In the limit $b\rightarrow \infty$ we obtain linear
Maxwell relations: ${\bf D}={\bf E}$ and ${\bf H}={\bf B}$ (we use the
Heaviside-Lorentz system of units).

Now,  consider a point-like, Born-Infeld  charged particle at rest.  It is
described by the static solution of the Born-Infeld field equations
(\ref{Max1})--(\ref{H}) with $\rho=e\delta({\bf r})$ and ${\bf j}=0$,
where $e$ denotes the particle's electric charge.  Obviously ${\bf B}={\bf
H}=0$ (Born-Infeld electrostatics). Moreover, the spherically symmetric
solution of $\nabla\cdot{\bf D}=e\delta({\bf r})$ is given by the Coulomb
formula
\begin{eqnarray}  \label{D0}
{\bf D}_0 = \frac{e}{4\pi r^3}{\bf r}\ .
\end{eqnarray}
Using (\ref{D}) one easily finds the corresponding ${\bf E}_0$ field
\begin{eqnarray}    \label{E0}
{\bf E}_0 = \frac{{\bf D}_0}{\sqrt{1+b^{-2}{\bf D}_0^2}} =
\frac{e}{4\pi r}\frac{{\bf r}}{\sqrt{r^4+r_0^4}}\ ,
\end{eqnarray}
where we introduced
$r_0:= \sqrt{\frac{|e|}{4\pi b}}$.
Note, that the field ${\bf E}_0$, contrary to ${\bf D}_0$, is bounded in
the vicinity of a particle: $|{\bf E}_0|\leq \frac{|e|}{4\pi r_0^2} = b$.
It implies that the energy of a point charge is already finite.

Now, let us perturb the static Born-Infeld solution $\bE_0$:
\begin{equation}   \label{perturb-E}
\bE = \bE_0 + \tE\ ,
\end{equation}
where $\tE$ denotes a weak perturbation, i.~e. $|\tE|\ll |\bE_0|$.  The
corresponding $\bD$ field may be obtained from (\ref{D}). In the
electrostatic case, i.~e. ${\bf B}={\bf H}=0$, formula (\ref{D}) reads:
\begin{equation}    \label{E-D}
\bD = \frac{\bE}{\sqrt{1-b^{-2}{\bE}^2}}\ ,\ \ \ \ \  \
\bE = \frac{\bD}{\sqrt{1+b^{-2}{\bD}^2}}\ .
\end{equation}
Therefore, using (\ref{perturb-E}) we get
\begin{equation}   \label{bD}
\bD =  \bD_0 +
\frac{1}{\sqrt{1-b^{-2}\bE_0^2}} \left( \tE + b^{-2}(\bD_0\tE)\bD_0 \right)
 + O(\tE^2) \ ,
\end{equation}
where $O(\tE^2)$ denotes terms  vanishing for $|\tE|\rightarrow 0$ like
$\tE^2$ or faster. In the present paper we study only the {\it linear
perturbation}, i.~e.~we keep in (\ref{bD})  terms linear in $\tE$ and
neglect $O(\tE^2)$. Therefore, in this approximation the perturbation $\tD
= \bD - \bD_0$ of the field $\bD_0$ equals:
\begin{eqnarray}   \label{perturb-D}
\tD = \frac{1}{\sqrt{1-b^{-2}\bE_0^2}} \left( b^{-2}(\bD_0\tE)\bD_0
+ \tE \right) = \sqrt{1+b^{-2}\bD_0^2} \left( b^{-2}(\bD_0\tE)\bD_0 + \tE
\right)\ .
\end{eqnarray}
The fields $\tE$ and $\tD$ fulfill the following equations:
\begin{eqnarray}
\nabla \times \tE=0\ ,\ \ \ \ \
\nabla\cdot\tD=0\ .
\end{eqnarray}
The first one implies that $\tE=-\nabla\tph$. Therefore, using
(\ref{perturb-D}),  $\nabla\cdot\tD=0$ leads to the following equation for
the potential $\tph$:
\begin{eqnarray}  \label{tilde-phi}
\Delta\tph + \left( \frac{r_0}{r}\right)^4 \left[
  \frac{\partial^2
\tph}{\partial r^2}
-\frac4r   \frac{\partial\tph}{\partial r} \right] = 0\ ,
\end{eqnarray}
where $\Delta$ stands for a 3-dimensional Laplace operator in ${\bf R}^3$.
Observe, that in the limit $r_0\rightarrow 0$ (i.~e. Maxwell theory) we
obtain simply the Laplace equation $\Delta\tph=0$ for the electrostatic
potential $\tph$. Using spherical coordinates in ${\bf R}^3$ the Laplace
operator $\Delta$ reads:
\begin{equation}
\Delta \tph =    \frac{1}{r}
\frac{\partial^2}{\partial r^2} (r\tph) + \frac{1}{r^2}\, {\bf L}^2\,\tph  \ ,
\end{equation}
where ${\bf L}^2$ denotes the Laplace-Beltrami operator on the unit sphere
(it is equal to the square of the quantum-mechanical angular momentum).

We see, that due to the spherical symmetry of the unperturbed solution
$\bD_0$, different multipole modes decouple in the above equation.
Therefore, any solution of (\ref{tilde-phi}) may be written as follows:
\begin{eqnarray}         \label{sum}
\tph({\bf x}) = \sum_{l=1}^{\infty} a_l \tph_l({\bf x})\ ,
\end{eqnarray}
where
\begin{equation}             \label{tph}
\tph_l(r,\mbox{angles}) := \frac{\Psi_{{l}}(r)}{r}\ Y_l(\mbox{angles})\ ,
\end{equation}
and $Y_l$ denotes the $l$-pole eigenfunction of ${\bf L}^2$, i.~e.  ${\bf
L}^2 Y_l = -l(l+1)Y_l$. Obviously, an eigenfuction $Y_l$ is related to the
$l$-pole moment tensor by
\begin{eqnarray}  \label{eigen}
Y_l = r^{-l}\ x^{i_1}...x^{i_l}\, {\cal Q}_{i_1i_2...i_l}\ .
\end{eqnarray}

Observe, that we do not consider the monopole term
(i.~e.
$a_0\,\tph_0$) in (\ref{sum}). This term
corresponds to the gauge transformation of $\tph$ and due to the
gauge invariance of (\ref{tilde-phi}) it is unessential (we may fix the
gauge by putting e.~g.
$a_0=0$).

Let us look for  the  $l$-pole-like  deformation, i.~e. for a function
$\tph_l$.  Inserting the  {\it ansatz} (\ref{tph}) to (\ref{tilde-phi}) we
obtain the following equation for $\Psi_{{l}}(r)$:
\begin{equation}   \label{Psi}
\left(\Psi_{{l}}'' - \frac{l(l+1)}{r^2}\Psi_{{l}}\right) +
\left(\frac{r_0}{r}\right)^4
\left( \Psi_{{l}}'' - \frac 6r \Psi_{{l}}' + \frac{6}{r^2} \Psi_{{l}}
\right) = 0\ ,
\end{equation}
where $\Psi_{{l}}'$ stands for $\partial\Psi_{{l}}/\partial r$.  In the next
section we find 2-dimensional space of solution of (\ref{Psi}).

\section{Exact solution of the ``deformed'' Laplace equation}
\label{solution}

Let us note that
for $r \gg r_0$, equation (\ref{tilde-phi}) reduces to the standard
Laplace equation with 2 independent $l$-pole-like solutions: the one
corresponding to the constant $l$-pole moment (it behaves like $r^{l+1}$)
and the external $l$-pole solution (behaving like $r^{-l}$).

On the other hand, a basis may be chosen corresponding to the behaviour of
$\Psi_{{l}}$ at $r \rightarrow 0$. From the asymptotic analysis of (\ref{Psi})
it follows that
there is a solution
which behaves like $r^6$ and another one which
behaves like $r$.
Let us define
$\Phi_{l} := r^{-6}\Psi_{l}$ and introduce the following variable:
\begin{equation}   \label{z}
z:=-\left(\frac{r}{r_0}\right)^4\ .
\end{equation}
Using (\ref{Psi}) one gets the following equation for $\Phi_{l}$:
\begin{equation}   \label{Phi}
z(1-z) \frac{d^2\Phi_{{l} }}{dz^2} + \left(\frac 94 - \frac{15}{4}
z\right)
\frac{d\Phi_{{l}}}{dz} - \frac{30-l(l+1)}{16} \Phi_{{l}} = 0
\end{equation}
This is a hypergeometric equation (cf. \cite{Abramowitz}).
A hypergeometric equation for a function $u=u(z)$
\begin{eqnarray}  \label{hyper}
z(1-z)\,u'' + [c -(a+b+1)]\,u' - ab\,u =  0\ ,
\end{eqnarray}
 has two independent
solutions. One of them is given by the hypergeometric function
$_2F_1(a,b,c,z)$.  The other one has the following form (for  $c\neq 1$):
\begin{eqnarray}
z^{1-c}\, _2F_1(b-c+1,a-c+1,2-c,z)\ .
\end{eqnarray}
Therefore, the general solution of (\ref{Psi}) reads:
\begin{eqnarray}   \label{Psi-l}
\Psi_l(r) &=& A_l\, r^6\,  _2F_1\left(\frac{l+6}{4},\frac{5-l}{4},\frac
94,-\left(\frac{r}{r_0}\right)^4\right) \nonumber\\ &+&
B_l\, r\, _2F_1\left(\frac{l+1}{4},-\frac{l}{4},-\frac 14,
-\left(\frac{r}{r_0}\right)^4\right)\ .
\end{eqnarray}
The first term on the r.h.s. of (\ref{Psi-l}) behaves at $r\rightarrow 0 $
like $r^6$, the other one like $r$.

Let us note, however, that  the second term (behaving like $r$) corresponds to
the unphysical
solution. To see this let us look for the behaviour of the electric field
$\tE$ ``produced'' by it  in the vicinity of the Born-Infeld
particle, i.~e. for $r\rightarrow 0$. From (\ref{tph}) and (\ref{eigen}) we
get
\begin{eqnarray}    \label{singular-E}
\widetilde{E}_k = - \partial_k\tph &=&
- \frac lr [x \cdot {\cal Q}]_m \left( \delta^m_{\ k} -
\frac{x^mx_k}{r^2} \right) \nonumber\\ &+&
 l(l+1) \frac{r^3}{4r_0^4}
\left( \frac{4-l}{r} [x\cdot {\cal Q}] x_k  +
l [x \cdot {\cal Q}]_k \right) + O(r^7)\  ,
\end{eqnarray}
where
\begin{eqnarray}
[x\cdot {\cal Q}] &:=& r^{-l}x^{i_1}...x^{i_l}{\cal Q}_{i_1...i_l}\ ,\\ {[x\cdot {\cal Q}]}_k
&:=& r^{-l+1}x^{i_1}...x^{i_{l-1}}{\cal Q}_{i_1...i_{l-1}k}\ .
\end{eqnarray}
Therefore, due to the first term in (\ref{singular-E}), $\tE$  exceeds the
unperturbed field $\bE_0$ itself.

Moreover, the ``perturbation'' $\tE$ leads to infinite variation of the
total field energy. The ``electrostatic'' energy ${\cal H}$ corresponding
to electric induction {\bf D} is given by (cf. \cite{BI}, \cite{IBB}):
\begin{eqnarray}
{\cal H} = \int b^2\left( \sqrt{1+b^{-2}{\bf D}^2} - 1\right)d^3x\ .
\end{eqnarray}
Therefore, its variation reads
\begin{eqnarray}
\delta {\cal H}|_{\bD_0} = \int  \frac{D^k_0\delta D_k}{\sqrt{1+b^{-2}{\bf
D}_0^2}} \,d^3x\ = \int   E_0^k\delta D_k\, d^3x =  \int   E_0^k
\widetilde{D}_k\,  d^3x\ .
\end{eqnarray}
Now, let us investigate the behaviour of $\bE_0\tD$ for $r\rightarrow 0$.
From (\ref{perturb-D}) it follows that this expression contains highly
singular term which behaves like $r^{-6}\times \mbox{radial\ component\
of}\ \tE$. Using the expansion (\ref{singular-E}) we see that the first
term is purely tangential, i.~e. it is orthogonal to any  radial vector.
However, the second one does contain a radial part
$(e/4\pi)r^3l(l+1)[x\cdot {\cal Q}]$ and this way $\bE_0\tD$ produces
nonintegrable singularity $\sim r^{-3}[x\cdot {\cal Q}]$.  Therefore, we conclude
that the solution behaving like $r$ for $r\rightarrow 0$ has to be excluded
from our consideration, i.~e. $B_l=0$.
This way the solution of (\ref{Psi}) is given by
\begin{eqnarray}   \label{Psi-solution}
\Psi_l(r) = A_l\, r^6\,  _2F_1\left(\frac{l+6}{4},\frac{5-l}{4},\frac
94,-\left(\frac{r}{r_0}\right)^4\right) \ .
\end{eqnarray}

\section{Multipole moments of the particle}
\label{moments}

Knowing the exact solution of (\ref{Psi}) we are ready to calculate 
the $l$-th deformability coefficient of the Born-Infeld particle. 
We know that  at infinity, i.~e. for $r \gg r_0$, $\Psi_l$ is a 
combination of 2 solutions: one behaving like $r^{l+1}$ 
(it corresponds to the constant $l$-pole field) and
the other one behaving like $r^{-l}$ (corresponding to the external 
$l$-pole solution). Due to the fact that the space of physically 
admissible solution is only  1-dimensional (see
 (\ref{Psi-solution})) the proportion between these two solutions 
is no longer arbitrary. The ratio between them has therefore physical 
meaning. To find this ratio we shall use
the following property of a hypergeometric function $_2F_1(a,b,c,z)$ (cf.
\cite{Abramowitz}):
\begin{eqnarray}   \label{identity}
_2F_1(a,b,c,z) &=& \frac{ \Gamma(c)\Gamma(b-a) }{ \Gamma(b)\Gamma(c-a) }
(-z)^{-a}\, _2F_1\left(a,a+1-c,a+1-b,\frac 1z \right) + \nonumber \\
 &+& \frac{ \Gamma(c)\Gamma(a-b) }{ \Gamma(a)\Gamma(c-b) }
(-z)^{-b}\, _2F_1\left(b,b+1-c,b+1-a,\frac 1z \right) \ ,
\end{eqnarray}
where $\Gamma$ denotes the Euler $\Gamma$-function. This identity allows us
to find the asymptotic behaviour of $\Psi_l$ at infinity. Using
(\ref{Psi-solution}) and (\ref{identity}) one immediately
gets:
\begin{eqnarray}  \label{asymptotics}
\Psi_l(r)      &=& X_l\,
r^{l+1}\,  _2F_1\left(-\frac l4,\frac{5-l}{4},\frac{3-2l}{4},
-\left(\frac{r_0}{r}\right)^4\right) +   \nonumber\\ &+&
Y_l\,
r^{-l}\, _2F_1\left(\frac{l+6}{4},\frac{l+1}{4}, \frac{5+2l}{4},
-\left(\frac{r_0}{r}\right)^4\right)\ ,
\end{eqnarray}
where
\begin{eqnarray}                \label{Xl}
X_l &=& A_l\ \frac{ \Gamma(\frac{9}{4})\Gamma(-\frac{2l+1}{4}) }
            { \Gamma(\frac{l+6}{4})\Gamma(\frac{l+4}{4}) }\ r_0^{5-l}\ ,\\
\label{Yl}
Y_l &=& A_l\ \frac{ \Gamma(\frac{9}{4})\Gamma(\frac{2l+1}{4}) }
            { \Gamma(\frac{5-l}{4})\Gamma(\frac{3-l}{4}) }\ r_0^{6+l}\ .
\end{eqnarray}
Note, that the first term on the r.h.s. of (\ref{asymptotics}) behaves at
infinity like
$r^{l+1}$ and corresponds to the constant $l$-pole field 
${{\bf E}}_{\infty}$ given by the formula (\ref{E-infty}). Te second one
behaving like $r^{-l}$ corresponds to the external $l$-pole solution
${{\bf E}}_{\cal M}$ given by (\ref{E-M}). We interpret the
ratio between these two ingredients in formula (\ref{asymptotics})
\begin{eqnarray}
\kappa_l := \frac{Y_l}{X_l}\ ,
\end{eqnarray}
as the ``deformability coefficient'' (more precisely, the $l$-th
deformability coefficient) of the Born-Infeld particle.
 It measures the particle's
$l$-pole
moment ${\cal M}_{i_1...i_l}$ generated by the constant $l$-pole
moment ${\cal Q}_{i_1...i_l}$ of
the external electric field. Using (\ref{Xl}) and (\ref{Yl}) we obtain:
\begin{eqnarray}   \label{kappa-l}
\kappa_l = \frac{
\Gamma(\frac{l+6}{4})\Gamma(\frac{l+4}{4})\Gamma(-\frac{2l+1}{4})}
  { \Gamma(\frac{5-l}{4})\Gamma(\frac{3-l}{4})
  \Gamma(\frac{2l+1}{4})}\ r_0^{2l+1}\ .
\end{eqnarray}
Obviously, in the limit of Maxwell theory ($r_0\rightarrow 0$), the
external electric field does not generate any particle's $l$-pole moment.
Therefore, this mechanism comes entirely from the non-linearity of the
Born-Infeld theory.

\section{Physical implications}

We used in this paper very specific model of non-linear electrodynamics.
There is a natural question: why this particular  model?
It turns out that among other non-linear theories of electromagnetism,
Born-Infeld theory possesses very distinguished physical properties
\cite{Hagiwara}. For example it is the only causal spin-1 theory 
\cite{Plebanski} (aside from the Maxwell theory). The assumption, that
the theory is effectively non-linear in the vicinity of a charged particle
is very natural from the physical point of view. Actually, we have learned
this from quantum electrodynamics.
 There were attempts to identify non-polynomial
 Born-Infeld  lagrangian as an effective Euler-Heisenberg lagrangian
\cite{Euler}. It was shown \cite{Hagiwara1} that the effective 
lagrangian can coincide with the (\ref{Lag-BI}) up to six-photon
interaction terms. Recently, there is a new interest in the Born-Infeld
electrodynamics due to the investigation in the string theory (see e.~g. 
\cite{string}), where (\ref{Lag-BI}) is not postulated but derived.

Now, let us  make some comments concerning physical implications 
of the obtained results. First of all taking $l=1$ one gets
\begin{eqnarray}  \label{kappa1}
\kappa_1 = -1.85407\  r_0^3\ ,
\end{eqnarray}
which reproduces result obtained in \cite{dipole}.
In \cite{dipole} it was mentioned that for $l=1$
one might describe, this way, the polarizability coefficient of the
proton. According to \cite{Bar}, we have $\kappa_1 = (12.1 \pm 0.9)
\times 10^{-4}\ \mbox{fm}^3 $. To fit this value one has to take
$r_0 \approx 0.09\ \mbox{fm}$. The total mass of the corresponding Born-Infeld
unperturbed field accompanying such a particle is about 32 electron
masses. We see, that the main part of the proton total mass cannot be
of electromagnetic nature and has to be concentrated in the material
core of the particle.

Unfortunatelly, there are no experimental data concerning particles
polarizability for $l\neq 1$. Probably, it is highly nontrivial task
to measure these
quantities experimentally. It  would be very interesting to have the
possibility to compare such quantities with the formula (\ref{kappa-l}).

Let us note that $\kappa_3$ and $\kappa_5$ vanish due to $\Gamma(0)$
in the denominator of (\ref{kappa-l}). All other $\kappa_l \neq 0$. It does
not
mean that the particle is not polarizable for $l=3$ and $l=5$. It is true in
the linear  approximation only. However, this result suggests that in these
two
sectors it is much more difficult to polarize the particle than in the other
ones.

\section*{Acknowledgement}

D.C. thanks Alexander von Humboldt Stiftung for 
the financial support.

\end{document}